# Introduction

The introduction of water vapor in a plasma source may lead to some undesired chemicals reactions [1] and instabilities [2]. However, water vapor can also be deliberately mixed with the plasma gas to generate radicals of interest such as OH. This approach has already been investigated in different domains: surface treatments [3], biocompatibility [4] and plasma medicine [5]. In surface treatments, the oxidative functionalization can also be achieved with $O_2$ [6] but promoting the mixture of water vapor with the carrier gas induces a milder treatment [7].

# 2. Experimental setup

The study of the water reactivity in plasma was achieved by injecting water vapor in a dielectric barrier discharge (DBD) operating at atmospheric pressure as shown in Fig. 1. The system consists in two circular (8 cm of diameter) copper electrodes recovered with a dielectric. These two dielectrics were made in alumina and their thickness was 2 mm. The two electrodes were separated by a distance of 5 mm, the upper electrode was AC biased whereas the lower electrode was grounded. Argon was used as carrier gas with a flow fixed at 15 L/min, and the plasma was powered at 20 W. The water was maintained at room temperature (18 °C) in a bubbler and injected into the discharge for flow rates comprised between 0 and 2.6 mL/s.

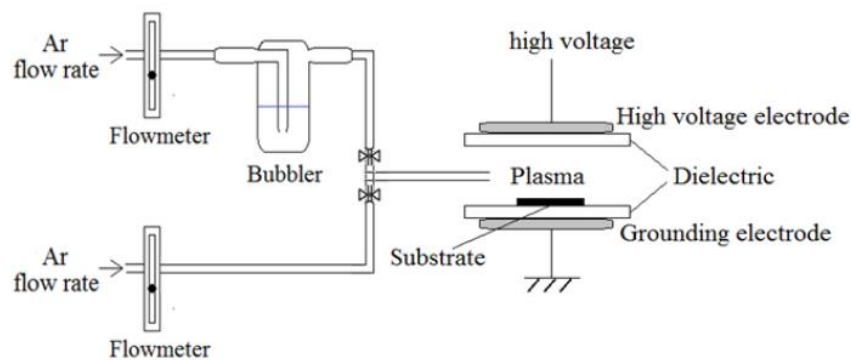

*Fig. 1. Schematic diagram of the experimental setup.*

Different techniques were used to characterize the treated LDPE. A drop shape analyzer (Krüss DSA 100) was employed to measure static contact angles of water drops deposited onto LDPE samples. Drops of 3 µL were deposited on the surface and an average of 10 different drops was used as the final measure.

To evaluate the chemical composition at the surface of the samples, XPS analyzes were performed with a Physical Electronics PHI-5600 instrument. The pressure in the analytical chamber was ≈$10^{-9}$ mbar. Survey scans were used to determine the chemical elements present at the surface such as O (1s). Spectra were acquired using the Mg anode (1253.6 eV) operating at 300 W. Wide surveys were acquired at 93.9 eV pass-energy, with a 10 scans accumulation (time/step: 50 ms, eV/step: 0.8).

The surface morphology was further analyzed with atomic force microscopy. AFM images were recorded in air with a Nanoscope IIIa microscope operated in tapping mode. The probes were silicon tips with a spring constant of 24-52 N.m$^{-1}$, a resonance frequency lying in the 264-339 kHz range and a typical radius of curvature in the 5-10 nm range. The images presented were height (5 µm × 5 µm).

Optical emission spectroscopy (OES) has been performed with a Spectra Pro-2500i spectrometer from ACTON research Corporation (0.500 m focal length, triple grating imaging). The light emitted was collected by an optical fiber and then was transmitted to the entrance slit (50 µm) of the monochromator where it was collimated, diffracted, focused on the exit slit and finally captured by a CCD camera from Princeton Instruments. Each optical emission spectrum was acquired with a 1800 grooves mm$^{-1}$ grating (blazed at 500 nm) and recorded on 30 accumulations with an exposure time of 25 ms. For every $H_2O$ flow rate, the emissions of all the species have been divided by the total emission of the discharge (i.e., a continuum from 250 to 850 nm).

## 3. Results

Production and consumption rates of Ar, Ar metastable, atomic O and OH were evidenced by optical emission spectroscopy (OES). The influence of the water vapor flow rate on these species (Fig. 2) was highlighted to have a better understanding of the reactivity. Some chemical reactions occurring within the discharge could be evidenced and their importance could be weighted according to their kinetic constants, more specifically, in the case of the OH radicals.

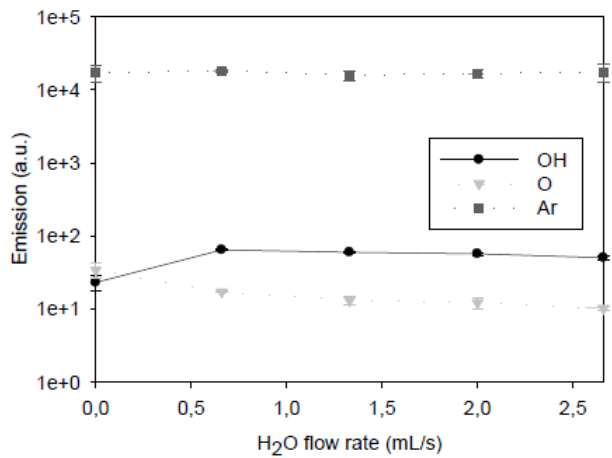

*Fig. 2. Emission of OH, O and Ar species versus the water flow rate in the Ar/water DBD.*

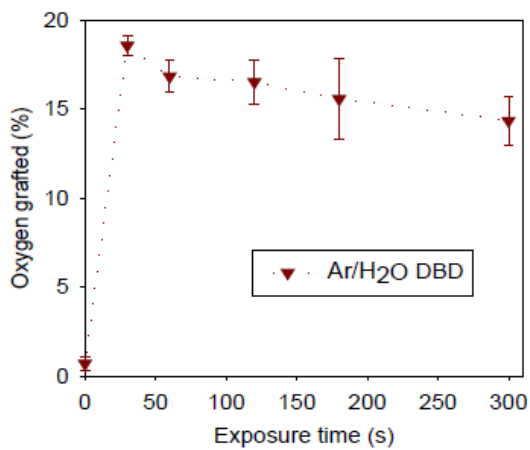
Fig. 3. Evolution of the oxygen grafted on LDPE surface.

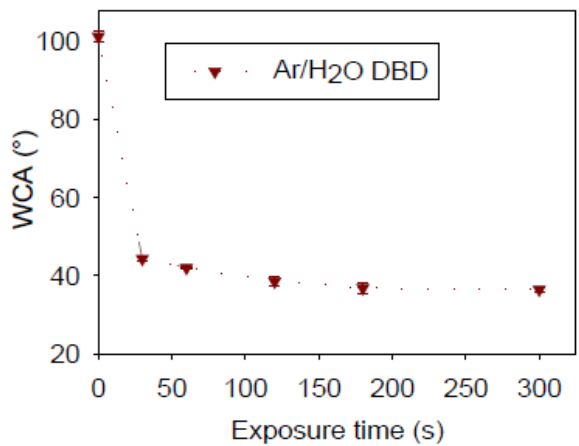
Fig. 4. Increase in hydrophilicity determined by WCA.

Indeed, an increase of OH was observed when water was injected into the DBD. That can be explained by these reactions (Table 1) leading to a partial dissociation of $H_2O$ and then a production of OH radicals. The trend between the increase of OH and the decrease of O can be linked. From 0 to 0.7 mL/s, the consumption of atomic O was counterbalanced by the production of OH radicals. Beyond 0.7 mL/s, an equilibrium seems to be reached for the consumption of O radicals, hence a limitation in the emission of the OH radicals.

| Reaction | Rate constant | Ref. |
|---|---|---|
| $Ar^M + H_2O \rightarrow Ar + OH + H$ | $k = 4.5 \cdot 10^{-10}$ | [8] |
| $O(^1D) + H_2O \rightarrow OH + OH$ | $k = 2.2 \cdot 10^{-10}$ | [8] |
| $O(^3P) + H_2O \rightarrow OH + OH$ | $k = 2.5 \cdot 10^{-14}$ | [8] |
| $OH + OH \rightarrow H_2O + O$ | $k = 2.5 \cdot 10^{-15}$ | [8] |

Table. 1. Main possible reactions occurring in the DBD.

The reactivity of $H_2O$ in the discharge was also carried out using an indirect method: the exposure of low density polyethylene (LDPE) samples to the plasma, in order to correlate the amount of oxygenated radicals resulting from water vapor dissociation reactions with the amount of oxygenated functions (C-O, C=O, COO) grafted on the surface. The XPS measurements achieved on the O 1s peak reveal a strong increase in the oxygen present at the surface for the first 35 s of treatment. An increase from almost 0 to 18% was observed (Fig. 3). After 35 s, the oxygen concentration slowly decays until a value of 15% reached after 300 s. These measurements have been correlated with WCA measurements expressed as a function of time (Fig. 4).

Between 0 and 35 s, a strong decrease in the WCA was observed from 100° to 43° and can be linked to the rise of the O%. However, at time higher than 35 s, the WCA still decreases but less significantly. AFM measurements were performed to observe a possible texturization of the treated LDPE. Fig. 5 shows an increase of the roughness of the surface with the treatment time. Indeed, after 60 s of treatment, a competitive effect can be assumed and can be explained by the Wenzel equation [9]:

$$\cos \Theta_{apparent} = r \cos \Theta$$

where r is the roughness, $\Theta_{apparent}$ is the apparent contact angle at the stable equilibrium state, $\Theta$ is the Young contact angle as defined for an ideal surface. Indeed, the "r" factor in the Wenzel corresponds to the roughness which was measured by AFM and increases with time. $\cos\Theta$ can be linked to oxygen percentage obtained by XPS. The percentage of O decreased when the time increases leading to a decrease in the cos. These two factors have opposite effect which could be compensated and is in agreement with the WCA measurement that gives information on the apparent cos Θ which remains constant.

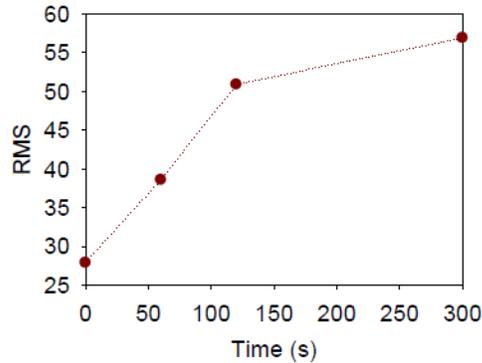

*Fig. 5. Increase in the roughness of the LDPE surface measured by AFM.*

## 4. Acknowledgment

This work was supported by PSI-IAP 7: plasma surface interactions (Belgian Federal Government BELSPO agency).

## 5. References

[1] N. Srivastava and C. Wang. J. Appl. Phys., 110 (2011)

[2] I. Koo and W. Lee. Plasma Chem. Plasma Process., 24 (2004)

[3] V. Rybkin, E. Kuvaldina, A. Grinevich, A. Choukourov, H. Iwai and H. Biederman. Plasma Process. Polymers, 5 (2008)

[4] K.S. Siow, L. Britcher, S. Kumar and H.J. Griesser. Plasma Process. Polymers, 3 (2006)

[5] G. Fridman, G. Friedman, A. Gustol, A.B. Shekhter, V.N. Vasilets and A. Fridman. Plasma Process. Polymers, 5 (2008)

[6] T. Dufour, J. Minnebo, S. Abou Rich, E.C. Neyts, A. Bogaerts and F. Reniers. J. Phys. D: Appl. Phys., 47 (2014)

[7] L. Liu, D. Xie, M. Wu, X. Yang, Z. Xu, W. Wang, X. Bai and E. Wang. Carbon, 50 (2012)

[8] L. Magne, S. Pasquiers, K. Gadonna, P. Jeanney, N. Blin-Simiand, F. Jorand and C. Postel. J. Phys. D: Appl. Phys., 42 (2009)

[9] G. Wolansky and A. Marmur. Colloids Surf. A: Physicochem. Engng. Aspects, 156 (1999)